\documentclass[sigconf,screen]{acmart}

\AtBeginDocument{%
  }

\copyrightyear{2026}
\acmYear{2026}
\setcopyright{cc}
\setcctype{by}
\acmConference[UMAP '26]{34th ACM Conference on User Modeling, Adaptation and Personalization}{June 08--11, 2026}{Gothenburg, Sweden}
\acmBooktitle{34th ACM Conference on User Modeling, Adaptation and Personalization (UMAP '26), June 08--11, 2026, Gothenburg, Sweden}
\acmDOI{10.1145/3774935.3812707}
\acmISBN{979-8-4007-2311-7/2026/06}

\usepackage{xcolor}
\usepackage{cleveref}

\graphicspath{{./img/}}

\begin{document}

\title{Measuring the stability and plasticity of recommender systems}

\author{Maria João Lavoura}
\orcid{0000-0002-3718-4451}
\affiliation{%
  \institution{Independent researcher}
  \city{Lisbon}
  \country{Portugal}
}

\author{Robert Jungnickel}
\authornote{Corresponding author}
\orcid{0000-0002-9102-722X}
\affiliation{%
  \institution{Joint Research Centre - European Commission}
  \city{Seville}
  \country{Spain}
}

\author{João Vinagre}
\authornotemark[1]
\orcid{0000-0001-6219-3977}
\affiliation{%
  \institution{Joint Research Centre - European Commission}
  \city{Seville}
  \country{Spain}
}


\begin{abstract}
The typical offline protocol to evaluate recommendation algorithms is to collect a dataset of user-item interactions and then use a part of this dataset to train a model, and the remaining data to measure how closely the model recommendations match the observed user interactions. 
This protocol is straightforward, useful and practical, but it only provides snapshot performance. 
We know, however, that online systems evolve over time.  
In general, it is a good idea that models are frequently retrained with recent data. 
But if this is the case, to what extent can we trust previous evaluations? 
How will a model perform when a different pattern (re)emerges? 
In this paper we propose a methodology to study how recommendation models behave when they are retrained.
The idea is to profile algorithms according to their ability to, on the one hand, retain past patterns -- stability -- and, on the other hand, (quickly) adapt to changes -- plasticity. 
We devise an offline evaluation protocol that provides detail on the long-term behavior of models, and that is agnostic to datasets, algorithms and metrics. 
To illustrate the potential of this framework, we present preliminary results of three different types of algorithms on the GoodReads dataset that suggest different stability and plasticity profiles depending on the algorithmic technique, and a possible trade-off between stability and plasticity.
We further discuss the potential and limitations of the proposal and advance some possible improvements.  
\end{abstract}




\maketitle

\section{Introduction}
The stability-plasticity dilemma has been identified by \citet{mermillod2013stability} as a fundamental trade-off of learning systems. 
It consists of the tension between two potentially conflicting requirements: \emph{stability},  the ability to retain previously learned concepts, and \emph{plasticity}, the ability to learn new ones. 
A system with high plasticity will tend to be forgetful, whereas an overly stable system will struggle to adapt to changes.

Traditional offline evaluation protocols are agnostic to time, and therefore do not account for such properties. 
However, it is reasonable to hypothesize that different types of algorithms will learn models with disparate stability and plasticity profiles. 

The desirable stability and plasticity are dependent on the dynamics of the task at hand. 
Highly dynamic domains, such as news or social media, may favor models with high plasticity, while domains with few and slow changes, such as music or books, may benefit more from models with high stability.  

Besides algorithm selection, data management or cold-start issues can benefit from knowledge about the stability-plasticity profile of certain classes of models. Furthermore, regulatory requirements, such as those under the EU's Digital Services Act, require risk assessment and mitigation measures, that can be enabled by better predicting how systems will react to certain changes.
In short, evaluating stability and plasticity of recommendation models may have implications with high potential downstream impact.

In the field of continual learning \cite{DBLP:journals/pami/WangZSZ24}, researchers have looked at this problem, typically focusing on classic supervised machine learning in a sequence of well identified tasks -- the so-called task-incremental learning. 
The problem in recommendation is that such division does not exist naturally, so most of the methods to measure stability and plasticity used in previous research are not applicable to recommendation models. 
In addition, past work focuses disproportionally on the problem of catastrophic forgetting -- i.e. through the lens of stability --, with comparably very few works that focus on plasticity.

This paper proposes a framework that answers questions related to the ability of recommendation models to retain past concepts and assimilate new ones. 
Our research questions are the following:
\begin{itemize}
  \item RQ1: How can we measure and compare the stability and plasticity of different recommendation models using a common evaluation framework?
  \item RQ2: Do different algorithmic approaches exhibit distinct trade-offs?
\end{itemize}

Our two contributions are (1) an algorithm- and metric-agnostic methodology that can be used systematically to profile algorithms with respect to their stability-plasticity profile, and (2) preliminary insights on the stability-plasticity trade-offs of different classes of algorithms across metrics related to accuracy and diversity.

We structure the remainder of the paper with and overview of related work in \Cref{sec:soa}, a description of our methodology in \Cref{sec:methodology}, and results in \Cref{sec:experiments}. In \Cref{sec:conclusions} we present our conclusions, limitations and future work.

\section{Related work}
\label{sec:soa}

The concepts of stability and plasticity have been discussed for quite some time by researchers studying continual machine learning~\cite{DBLP:journals/pami/WangZSZ24}, as a broader framing of the problem of \emph{catastrophic forgetting}~\cite{french1999catastrophic}, suggesting there is a dilemma, or a trade-off, between the ability of intelligent systems to retain what they have learned from past tasks and their ability to learn new tasks~\cite{mermillod2013stability}.
%
In the related literature, a continual learning problem essentially consists of incrementally learning models over a sequence of similar machine learning tasks~\cite{DBLP:journals/corr/abs-1904-07734}. The challenge is not only to maximize performance on the most recently learned tasks, but also to minimize loss in previous ones. The main distinction between existing approaches is whether they require task identifiers in the data, or are able to infer tasks autonomously. 
%
One approach towards systematically measuring stability and plasticity is proposed by~\citet{DBLP:conf/nips/Lopez-PazR17} through the concepts of forward and backward information transfer. The two concepts are captured by metrics that reflect the ability of past models to perform on future tasks and future models to perform in pasts tasks, respectively. 

In recommender systems, the problem of continual learning has been addressed by~\citet{DBLP:conf/sigir/YooKQX0T25}, who propose a method to tackle the lack of plasticity. Stability and plasticity are not measured -- the authors focus instead on adding plasticity in the presence of concept drifts, similarly to the approach of stream-based recommender systems \cite{DBLP:journals/csur/Al-GhosseinAB21}, relying on the assumption that forgetting is beneficial in the presence of concept drifts~\cite{DBLP:journals/kais/MatuszykVSJG18}.

Our proposal draws inspiration from the above work, however a direct application of methods designed for classic machine learning such as \cite{DBLP:conf/nips/Lopez-PazR17} is not possible for multiple reasons. First, the idea of tasks is highly ambiguous in recommendation problems. In~\cite{DBLP:conf/sigir/YuanZKJKL21}, tasks are operationalized as the user activity in distinct platforms -- i.e. social media, marketplaces, etc. However, this requires a scenario where user representations are shared by all platforms. Second, unlike typical classification, the role of recommendation models is not to learn class distributions, which makes the direct application of existing methods inadequate for the nuance of recommendation problems. Finally, the variety of problem settings and evaluation protocols -- including e.g. splitting criteria and evaluation metrics -- complicate the application of methodologies that fundamentally rely on standard machine learning evaluation methods. 

\section{Methodology}
\label{sec:methodology}

Both stability and plasticity can only be observed in the presence of change.
A model trained on a stable dataset will be naturally stable, simply because the underlying distributions do not change significantly.
The problem is that we typically do not know if, when, and to what extend natural changes occur, unless we investigate deep into the dynamics of each dataset. 
Realistically, concept drifts and shifts, particularly in recommendation tasks, can be extremely complex given that they stem from an intricate combination of user behavior dynamics, changes in the system -- e.g. in the user interface --, and real-world exogenous variables.

We want a methodology that allows practitioners and system developers to profile and select models according to their needs. 
This methodology needs to be generally applicable and connected to realistic deployment scenarios for results to be meaningful.

In addition, we need metrics that effectively capture the concepts we are trying to measure, ideally agnostic to the accuracy (and beyond accuracy) measures we wish to focus on.

\subsection{Data dynamics}
\label{subsec:data_dynamics}

The most obvious way to study the reaction of a model to change is to introduce it artificially by manipulating the dataset.
This strategy has the advantage that we can control when, how and how much change is introduced. 
For the purpose of this paper -- and for the sake of simplicity --, we exclusively use simple user-item interaction data.
Our dataset $D$ consists of a sequence of user-item pairs $(u,i)$, which we assume to be  a positive preference of user $u$ for item $i$. 

Our strategy for introducing a shift follows previous work in \cite{vinagre2012}. 
We first split dataset $D$ into two equal time intervals, $D_1$ and $D_2$, respecting the order of interactions. For example, if D contains 2 years of data, each subset will consist of 1 year.
Then, we randomly sample 50\% of the items occurring in $D_2$ and simply change their labels -- or IDs -- to different new labels. 
We do this maintaining label consistency, which means that all selected instances of a given item $i$ occurring in $D_2$, become instances of a new item $i'$ in $D_2$.

These modified item labels will be seen by the training algorithm as entirely new items, forcing the model to change. 
However, by keeping 50\% of the original ones, we also limit the artificialization of the dataset, partially maintaining the natural patterns of interaction.

\subsection{Experimental Protocol}
\label{subsec:info_scheme}

Given a dataset $D$ with a sudden change as described above in \Cref{subsec:data_dynamics}, the strategy we follow is to simulate a plausible scenario in the real world, in which to have a model trained in a certain regime, and observing how this model performs in another regime. 
More specifically, we train two models $M_1$ and $M_2$:
\begin{itemize}
    \item $M_1$ is trained on $D_1$ only;
    \item $M_2$ is trained on both $D_1$ and $D_2$. 
\end{itemize}

Importantly, both $D_1$ and $D_2$ are split in training and test subsets $D_1^{Train}$, $D_1^{Test}$, $D_2^{Train}$ and $D_2^{Test}$. Both test sets $D_1^{Test}$ and $D_2^{Test}$ are withheld from all training tasks.
Model $M_1$ can be seen as a legacy model that has not seen the most recent set of data, and model $M_2$ would be a more recent model, after a typical retraining task, where the training algorithm is fed with past data together with more recent data. The fact that we know that $D_1$ and $D_2$ consist of different data regimes allows us to compare their performance.

\subsection{Metrics}
\label{subsec:metrics}

To measure \emph{plasticity}, we need to compare the performance on $D_2^{Test}$ of model $M_1$ that has not seen $D_2^{Train}$, and model $M_2$ that did. The hypothesis is that $M_2$ will naturally perform better than $M_1$ given that it has been trained in instances of $D_2$, while $M_1$ has not. The difference between the performance of the two models provides a good measure of how a certain model is able to adapt to the new regime in $D_2$.

With respect to \emph{stability}, a similar strategy is used, but we compare the performance of $M_1$ and $M_2$ on $D_1^{Test}$ instead. Knowing that $M_1$ has only seen instances of $D_1$ and $M_2$ has seen instances of both $D_1$ and $D_2$, we can hypothesize that $M_2$ will struggle with $D_1^{Test}$ because it has been exposed to a new interfering signal. The difference in performance of the two models will tell us how much is $M_2$ able to retain the patterns of $D_1$ after such interference.

\begin{figure}[!htb]
    \centering
    \includegraphics[width=0.7\columnwidth]{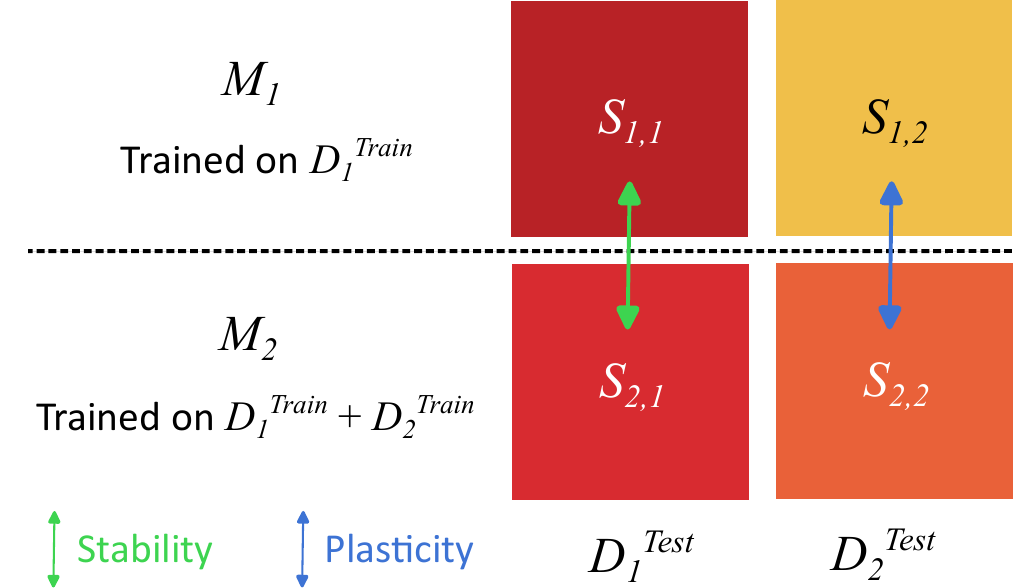}
    \caption{Experimental setup}
    \label{fig:benchmark_interpretation}
\end{figure}

\Cref{fig:benchmark_interpretation} illustrates this setup in a model vs test set matrix, as well as the comparisons (arrows) we associate to stability and plasticity. 
To compute stability and plasticity metrics, we collect four testing scores corresponding to the performance of models $M_1$ and $M_2$ on $D_1^{Test}$ and $D_2^{Test}$. We denote these scores as $S_{1,1}$ and $S_{1,2}$ corresponding to $M_1$ and $S_{2,1}$ and $S_{2,2}$ corresponding to $M_2$.

We formulate stability as \cref{eq:stability} and plasticity as \cref{eq:plasticity}. We further assume that scores are normalized to a $[0,1]$ scale and higher values mean higher stability or plasticity. Besides these two restrictions -- that we impose mostly for interpretability -- the measures are agnostic to the underlying performance metric. 

\begin{equation}
\label{eq:stability}
    Stability = 1 - (S_{1,1} - S_{2,1})
\end{equation}

\begin{equation}
\label{eq:plasticity}
    Plasticity = S_{2,2} - S_{1,2}
\end{equation}

\section{Experiments and preliminary results}
\label{sec:experiments}

To illustrate the potential of the framework proposed on \Cref{sec:methodology}, we run an experiment to measure the difference in stability and plasticity profiles of three algorithms implemented in RecBole~\cite{DBLP:conf/cikm/ZhaoMHLCPLLWTMF21}:

\begin{itemize}
    \item User-based KNN (UKNN);
    \item BPRMF~\cite{DBLP:conf/uai/RendleFGS09};
    \item NeuMF~\cite{DBLP:conf/www/HeLZNHC17}.
\end{itemize}

The algorithms are baseline representatives of three algorithmic paradigms: classic neighborhood-based methods, matrix factorization, and neural approaches. The idea is to have a glimpse on how algorithms following these different paradigms differ in terms of stability and plasticity.
For each algorithm, we apply our framework to accuracy, using HitRatio@20 as the underlying metric.

\subsection{Dataset}
\label{subsec:dataset}
In our experiment, we use a subset of the Goodreads reviews dataset\footnote{\url{https://cseweb.ucsd.edu/~jmcauley/datasets/goodreads.html}}, that consists of user reviews of books made by users between 1 July 2012 and 1 January 2015. We extract only positive interactions, by collecting only user-item pairs of reviews with a rating of 5 stars (in a 1 to 5 star scale), along with the timestamp of the review, from a sample of 2.000 randomly selected users. 

We reserve a subset $D_0$ with all interactions between 1 July 2012 and 31 December 2012 for the pre-training of models, to mitigate eventual cold-start effects. Then we split the remaining data in two subsets, corresponding to $D_1$ and $D_2$ in our methodology (see \Cref{sec:methodology}). We divided data in approximately equal time periods of 1 year, with $D_1$ corresponding to year of 2013 and $D_2$ to 2014.

In addition, we removed users with less than 2 interactions in any of the subsets. After this, the dataset contained a total of 81.105 interactions between 1.488 users and 81.105 books, covering a time period between 1 July 2012 and 1 January 2015.

For splitting between training and testing, we used the leave-one-out protocol, meaning that each holdout set $D_1^{Test}$ and $D_2^{Test}$ consists of one ground truth interaction per user.

\subsection{Results}
\label{subsec:results}

\begin{figure}[!htb]
    \centering
    \includegraphics[width=0.6\columnwidth]{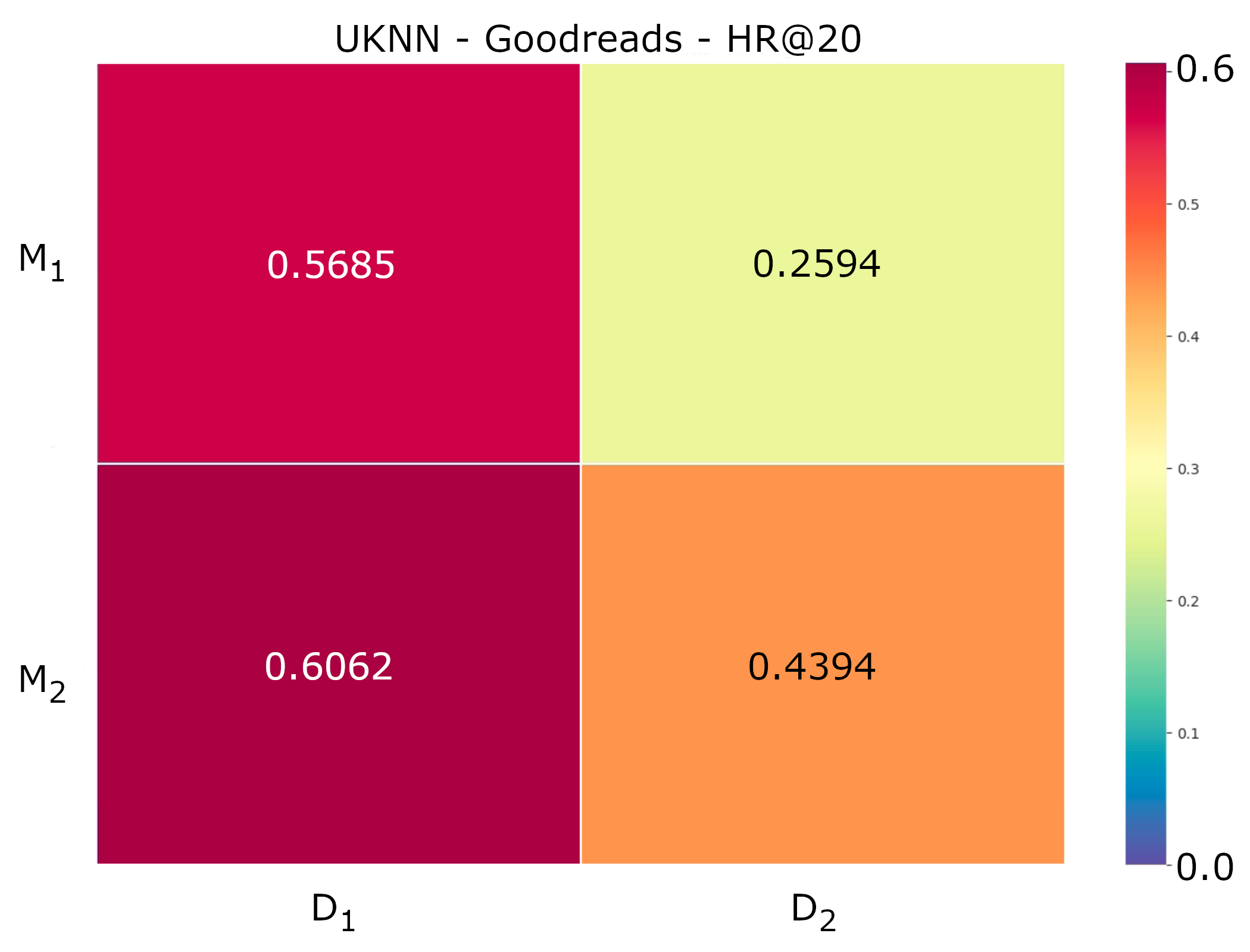}
    \caption{UserKNN model accuracy scores}
    \label{fig:UKNN}
\end{figure}
\begin{figure}[!htb]
    \centering
    \includegraphics[width=0.6\columnwidth]{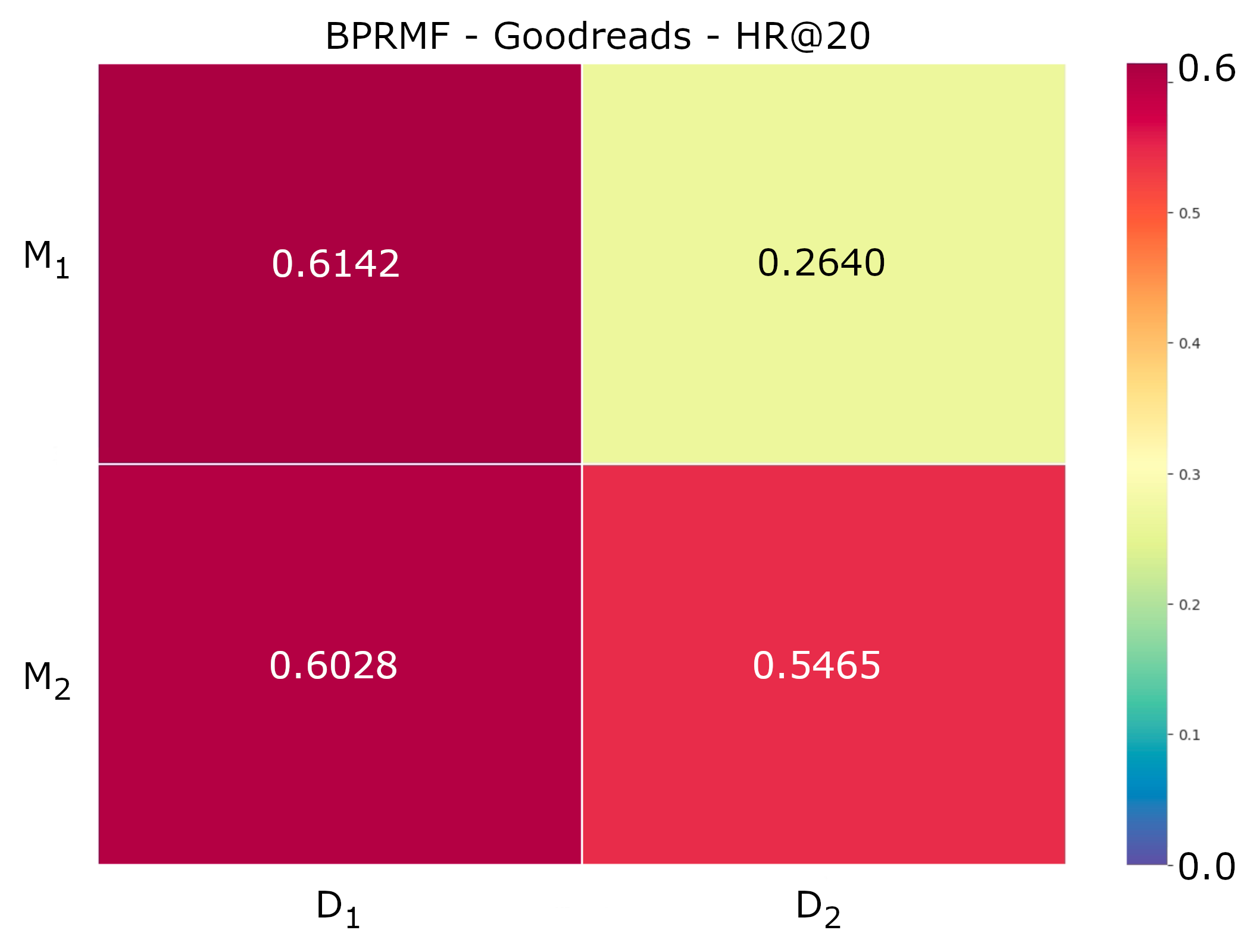}
    \caption{BPRMF model accuracy scores}
    \label{fig:BPR}
\end{figure}
\begin{figure}[!htb]
    \centering
    \includegraphics[width=0.6\columnwidth]{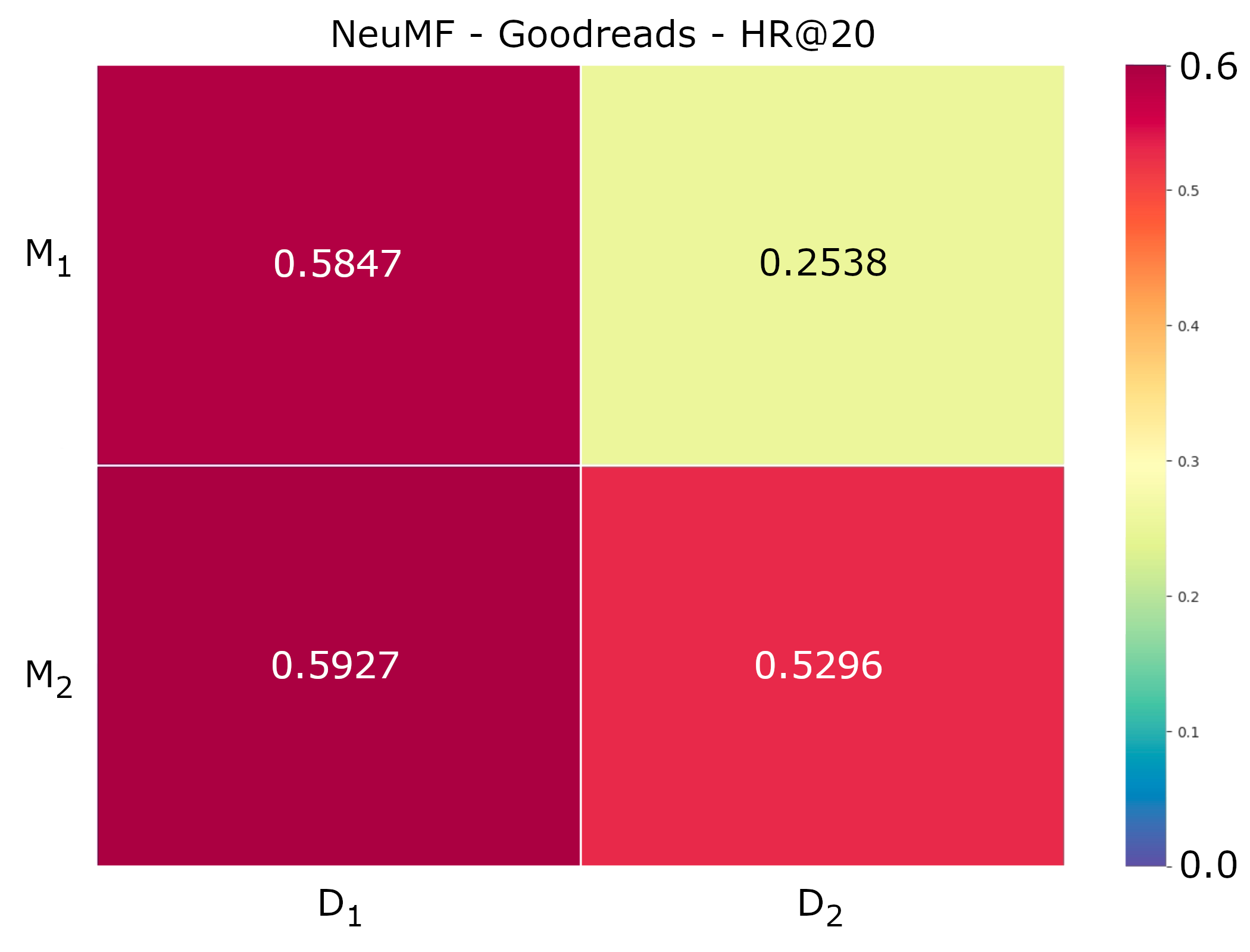}
    \caption{NeuMF model accuracy scores}
    \label{fig:NeuMF}
\end{figure}

Figures \ref{fig:UKNN}, \ref{fig:BPR} and \ref{fig:NeuMF} show the HitRatio@20 for each combination. By looking at the result heatmaps, we already spot some important differences between algorithms. For example, it is clear that UKNN has a stronger performance drop on $D_2^{Test}$ than the other two algorithms. Unsurprisingly, the worse performance is generally by $M_1$ on $D_2$, given the high proportion of items unseen by $M_1$.
By applying our stability and plasticity metrics to scores using \cref{eq:stability} and \cref{eq:plasticity}, we obtain the corresponding measures, in \Cref{tab:results}.

\begin{table}
    \centering
    \begin{tabular}{crr}
         Algorithm&  Stability& Plasticity\\
         \hline
         UKNN&  1.038& 0.180\\
         BPRMF&  0.989& 0.283\\
         NeuMF&  1.008& 0.276\\
    \end{tabular}
    \caption{Stability and Plasticity scores}
    \label{tab:results}
\end{table}

An interesting observation is that both UKNN and NeuMF actually improve the performance of $M_2$ with respect to $M_1$, leading to $Stability > 1$, which contradicts our intuition that interference of alien data in $D_2$ would generally affect the ability of $M_2$ to perform on $D_1^{Test}$. Regarding stability, results suggest that BPRMF is more unstable or, in other words, more susceptible to interference.
On the other hand, BPRMF seems to be more adaptable, since it has the highest score on $Plasticity$, which may indicate that there might be a trade-off between the two measures.
In a more pronounced way, we see a considerable gap in plasticity between UKNN and the two other models, suggesting that a KNN approach is more rigid and struggles to adapt to changes. On the other hand, it seems to keep and even considerably improve performance on past data.

Attributing differences to particular features or properties of the algorithms is beyond the scope of this work, however we can risk speculating that neighborhood-based models are more stable -- and less flexible -- given that they are essentially a form of memorizing user-item interactions. In other words, adding more data does not lead to the replacement of information. In contrast, model-based approaches, given their stochasticity and compactness, are necessarily more sensitive to some emerging patterns \emph{replacing} older ones. Experiments varying the size of neighborhoods of KNN approaches and the size of embeddings and model complexity in latent factor models might shed some light on what goes behind differences in algorithmic approaches.

\section{Conclusions, limitations and future work}
\label{sec:conclusions}
We present a framework to systematically measure the stability and plasticity of recommendation models. The framework is agnostic to the learning algorithms, dataset and evaluation metric of interest, and includes interpretable measures of both concepts. To illustrate the potential of the framework, we have applied it to the three distinct algorithms -- UKNN, BPRMF and NeuMF -- and used it to measure the stability and plasticity of the three algorithms on the Goodreads dataset, using an accuracy metric. Preliminary results allowed us to observe distinct stability-plasticity profiles. For example, it has allowed us to observe that UKNN has higher stability, but lower plasticity than latent factor models.
We are currently reviewing some of the assumptions behind our methodology, in particular the 50\% artificial shift introduced in the data which might be leading to a smaller impact than expected. In addition, the strategy to measure stability is not immune to temporal leakage -- $M_2$ has future knowledge w.r.t. $D_1$. We argue that this is unlikely to impact the relative stability of algorithms since they are all exposed to the same data, but recognize that it might introduce bias. We are currently working on an extended different version of the protocol that eliminates such bias. We are also working on further experiments with a larger set of datasets and algorithms, as well as metrics beyond accuracy, in particular coverage, diversity and fairness. Finally, we intend to publish a software package for easy experimentation and reproducibility.

\section*{Position statement}
The views expressed in this paper are those of the authors and  may  not,  under  any  circumstances,  be  regarded  as  an official position of the European Commission.


\bibliographystyle{ACM-Reference-Format}
\bibliography{references}



\end{document}